\documentclass{jpconf}

\usepackage[dvips]{graphicx}

\begin{document}

\title[Mott insulator to superfluid phase transition]{Dynamics of the Mott Insulator to Superfluid quantum phase transition in the truncated Wigner approximation}

\author{Marek Tylutki}
\ead{marek.tylutki@uj.edu.pl}
\address{
             Institute of Physics 
             and 
             Center for Complex Systems Research, 
             Jagiellonian University,
             Reymonta 4, 
             30-059 Krak\'ow, 
             Poland
}

\author{Jacek Dziarmaga}

\address{
             Institute of Physics 
             and 
             Center for Complex Systems Research, 
             Jagiellonian University,
             Reymonta 4, 
             30-059 Krak\'ow, 
             Poland
}

\author{Wojciech H. Zurek}

\address{Theory Division, Los Alamos National Laboratory, Los Alamos, NM 87545, USA}

\begin{abstract}
The quantum phase transition from the Mott insulator state to the
superfluid in the Bose-Hubbard model is investigated. We research one, two and three dimensional lattices in the truncated
Wigner approximation. We compute both kinetic and potential energy and
they turn out to have a power law behaviour as a function of the
transition rate, with the power equal to $1/3$. The same applies to the total
energy in a system with a harmonic trap, which is usually present
in the experimental set-up. These observations are in agreement with
the experiment of~\cite{deMarco1}, where such scalings were also observed and the power
of the decay was numerically close to $1/3$. The results confirm
the Kibble-Zurek (adiabatic-impulse-adiabatic approximation) scenario
for this transition.
\end{abstract}


\section{Introduction}

Excitations resulting from crossing the gapless quantum critical points are a serious problem for quantum simulations with ultracold atomic gases or ion traps. There, one would like to prepare a simple ground state of a simple initial Hamiltonian and then drive the system adiabatically to an interesting final ground state. This general observation has been recently supported by a more quantitative theory \cite{QuantumKZ,review}, that is by now confirmed by several numerical studies \cite{sabbatini}. The theory is a quantum generalization of the classical Kibble-Zurek mechanism (KZM) \cite{K,Z}. The theory predicts that density of excitations or excitation energy decay with a power of quench timescale $\tau_Q$. Experiments dedicated to the quantum theory were made in Refs. \cite{ferro,Mercedes,deMarco1}. In Ref. \cite{deMarco1} ultracold spinless bosonic atoms in a three-dimensional (3D) optical lattice were driven from the Mott insulator phase to the superfluid phase across a quantum phase transition. The transition was non-adiabatic and the excitation energy was reported to scale algebraically with the transition time, and the power was numerically equal to one third. Similarily, adiabaticity of loading atoms into an optical lattice \cite{Kuba} or releasing them from the lattice confinement \cite{deMarco2} has been questioned recently. The 1/3-scaling reported in the experiment \cite{deMarco1} coincides with the earlier prediction in Ref. \cite{Meisner} for 1D. In~\cite{dtz} it was shown, that the same scalings also hold for 3D. 

\section{The Bose-Hubbard model}

The Bose-Hubbard (BH) model is a lattice model that describes bosonic atoms in a field of an optical lattice~\cite{deMarco1,deMarco2,Kasevich,Greiner}. Its Hamiltonian assumes the following form
\begin{equation}
H_{\rm BH} = -J \sum_{\langle {\bf i},{\bf j} \rangle} \left( a_{\bf i}^\dag a_{\bf j} + a_{\bf j}^\dag a_{\bf i} \right)
  + \frac{1}{2n} \sum_{\bf i} a_{\bf i}^\dag a_{\bf i}^\dag a_{\bf i} a_{\bf i} + 
              \sum_{\bf i} V_{\bf i} a_{\bf i}^\dag a_{\bf i} ~,
\label{BH}
\end{equation}
where $a_{\bf i}$ is the annihilation operator at site ${\bf i}$ and $a^\dagger_{\bf i}$ is the corresponding creation operator. Here we do not specify the dimensionality of a lattice, which is three dimensional in real systems, but can be effectively made two- or one dimensional when some dimensions of the system are very small compared to the other. 

The BH Hamiltonian has three terms: (1) the kinetic term, which corresponds to the hopping of atoms between the neighbouring sites of the lattice, (2) the on-site interaction term, which is responsible for on-site repulsion between the atoms and possibly (3) the interaction with external potential, which could be a harmonic trap considered later in the article. Here, the coefficient $\frac{1}{2n}$ in front of the on-site interaction term is a result of our choice of units. 

When the kinetic term dominates over the on-site repulsion, the system is in the superfluid regime which is characterized by strong spatial correlations. On the other hand, when the on-site interaction term is much larger than the kinetic part, the system is in the Mott Insulator state, where there is a definite and constant number of atoms $n$ at each site,
\begin{equation}
| {\rm MI} \rangle = | n,n,n, \cdots \rangle ~.
\label{MI}
\end{equation} 

\begin{figure}[!ht]
\begin{center}
\includegraphics[width=.4\textwidth,clip=true]{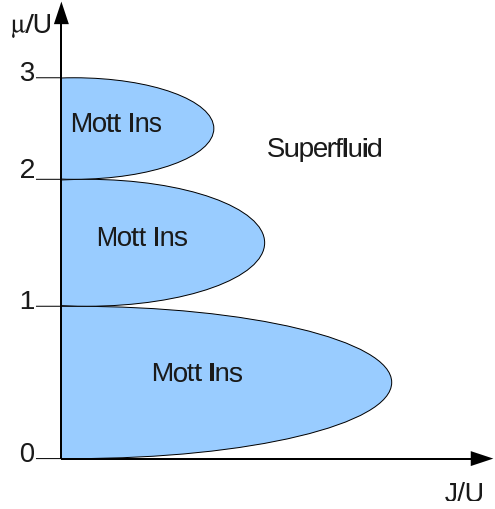}
\caption{The schematic phase diagram of the system defined by~(\ref{BH}). The lobes of the Mott insulator for $J \ll 1$ characterize with a constant number of atoms per site. The surrounding superfluid phase has a vanishing energy gap and non-zero correlations in phase between sites.}
\label{scheme}
\end{center}
\end{figure}

Clearly, between these two phases there is a boundary, crossing which causes the system to undergo a phase transition. The phase diagram of~(\ref{BH}) is depicted in Fig.~\ref{scheme}, where Mott Insulator phase forms the famous lobes, each characterized by a distinct value of atom density $n$. Those lobes are surrounded by a sea of the superfluid phase. 

The relative strength of the two terms is controlled by the coefficient $J$. To cross the phase boundary we vary this quantity in time. In this article, we assume that initially the system is in the Mott state and we drive it into the superfluid regime by increasing $J(t)$. This driven transition will be called here a {\it quench}, and we choose it to be linear in time:
\begin{equation}
J = \frac{t}{\tau_Q} ~,
\end{equation}
where $\tau_Q$ is the inverse of the quench rate. 

\section{The truncated Wigner method}

To investigate the model~(\ref{BH}), we adopt an approach known as the Truncated Wigner Method, where the full quantum dynamics of the original Hamiltonian is approximated by an evolution of a statistical ensemble of complex lattice fields $\phi_i$. This approximation is valid in the regime of large atom densities, $n \gg 1$, which we now assume. In the Truncated Wigner Method, we make an identification $a_i \approx \sqrt{n} \, \phi_i$, $a^\dagger_i \approx \sqrt{n} \, \phi^\star_i$. With this identification, the equation of motion for $a_i$ becomes the nonlinear equation for $\phi_i$
\begin{equation}
i\, \partial_t \phi_{\bf i} = 
-J \nabla^2\phi_{\bf i} + 
\left(|\phi_{\bf i}|^2-1\right)\phi_{\bf i}
\label{GPE} ~,
\end{equation}
which is known as the Gross-Pitaevskii equation (GPE). The quantum evolution of the original system~(\ref{BH}) is represented by the whole set of trajectories of $\phi_i$ and the quantum expectation values become statistical averages of relevant observables over the ensemble. The initial condition for the evolution becomes the distribution for $\phi_i$. The initial state~(\ref{MI}) translates to 
\begin{equation}
\phi_i(0) = e^{i \theta_i} ~,
\end{equation}
where $\theta$'s are uniformly distributed over $(-\pi,\pi]$. This reflects the uncertainty between the atom number and phase: $n$ is well defined and $\theta$ is completely random. 

The kinetic and potential energies take the form:
\begin{eqnarray}
E_{\rm kin} &=& J~\sum_{\bf i} \overline{ {\bf \nabla}\phi^*_{\bf i} {\bf \nabla}\phi_{\bf i} }~,\label{Ekin}\\
E_{\rm pot} &=& \frac12 \sum_{\bf i} \overline{\left(|\phi_{\bf i}|^2-1\right)^2} ~.\label{Epot}
\end{eqnarray}
The behaviour of energy with the rate of the transition will be at the center of our interest. 

\section{The impulse-adiabatic scenario}
\label{impAd}

\begin{figure}[!ht]
\begin{center}
\includegraphics[width=.6\textwidth,clip=true]{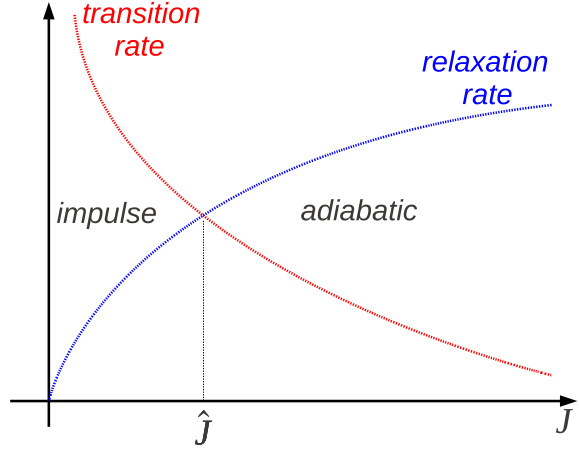}
\caption{The schematic depiction of the impulse-adiabatic approximation to the dynamics of the considered model. As long as the relative transition rate, $\dot{J}/J$, is larger than the relaxation rate, the dynamics is said to be impulse, i.e. the state of the system is frozen in spite of the changing Hamiltonian. When the relaxation rate begins to dominate, the state has more and more time to follow the Hamiltonian, and we assume that the evolution is adiabatic. }
\label{impulseAdiabatic}
\end{center}
\end{figure}

The next ingredient that we add to our analysis is the impulse-adiabatic scenario~\cite{Z}. It simplifies the picture of the evolution across the critical point. In the vicinity of that point we can treat the evolution as impulse, which means that the transition rate is much faster than the pace of the change of the system. Far from the critical point, on the contrary, the relaxation rate of the system exceeds the driving pace and the evolution is adiabatic. The idea of the impulse-adiabatic scenario is to say, that there is a certain value of the parameter, in our case $J = \hat{J}$ where the two rates become comparable, and which separates the impulse stage of the evolution from the adiabatic stage. In other words, below $\hat{J}$ the system undergoes the impulse evolution, and above - adiabatic. This idea is summarized in the Fig.~\ref{impulseAdiabatic}. 

In our units, the critical value of $J$ is $J_{\rm cr} \approx n^{-2}$ and tends to zero for large densities, $J \approx 0$. 

\section{The Josephson regime}

First, we restrict ourselves to the regime, where $J \ll 1$. In this regime the density fluctuations are small, $|\phi_{\bf i}|^2\approx 1$; the field amplitude does not deviate much from the Mott state's amplitude. Hence, it is convenient to parametrize 
\begin{equation}
\phi_{\bf i}~=~\left(1+f_{\bf i}\right)~e^{i\theta_{\bf i}} ~,
\label{ftheta}
\end{equation}
with real amplitude correction $f_{\bf i}$ and phase $\theta_{\bf i}$. This gives us equations of motion for $f_{\bf i}$ and $\theta_{\bf i}$, where the initial conditions are $f_{\bf i} = 0$ and random $\theta_{\bf i}$. After elimination of $f_{\bf i}\ll 1$ in Eq. (\ref{GPE}) we get the Josephson equations
\begin{equation}
\frac{d^2}{dt^2} \theta_{\bf i} =
2J
\sum_{{\bf j},\, {\rm n. n.}\, {\bf i}}
\sin\left(\theta_{{\bf j}}-\theta_{\bf i}\right)
\label{Josephson}
\end{equation}
with random initial $\theta_{\bf i}(0)$ and $\frac{d\theta_{\bf i}}{dt}(0)=0$. The summation in Eq.~(\ref{Josephson}) extends over the nearest neighbours of $\bf{i}$. 

The energies defined by Eqs.~(\ref{Ekin}, \ref{Epot}) in the approximation~(\ref{ftheta}) yield
\begin{eqnarray}\nonumber
E_{\rm kin} &\approx& J \sum_{\bf i} \overline{\nabla \theta_{\bf i} \cdot \nabla \theta_{\bf i}} ~,\\
E_{\rm pot} &\approx& 2 \sum_{\bf i} \overline{f_{\bf i}^2} ~.
\label{energyInJosephson}
\end{eqnarray}

\section{Thermalization}

\begin{figure}[!ht]
\begin{center}
\includegraphics[width=.9\textwidth,clip=true]{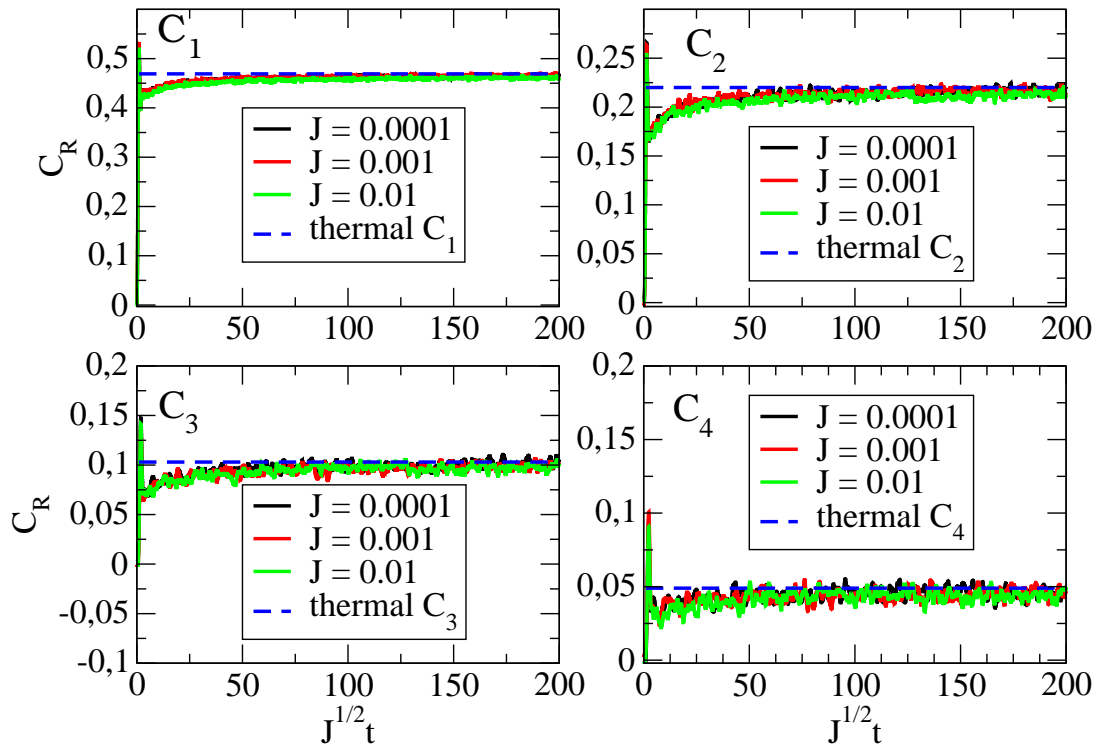}
\caption{Thermalization in 1D after a sudden quench, where $J$ abruptly jumps to its final value at $t = 0$. The short-range correlations quickly reach their equilibrium, whose numerical value was predicted in detail in~\cite{dtz}. In each panel, the plots for various $J$'s against time $J^{1/2} t$ collapse to a single curve, proving, that the thermalization time scales like $\tau \sim J^{-1/2}$}
\label{therm1D}
\end{center}
\end{figure}

\begin{figure}[!ht]
\begin{center}
\includegraphics[width=.6\textwidth,clip=true]{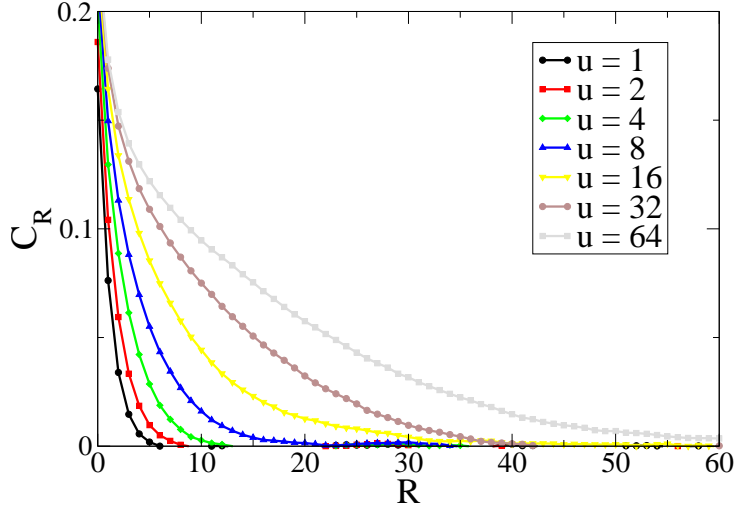}
\caption{Correlations in 3D. The correlation function $C_R = \overline{\phi^*_s \phi_{s+R}}$ for several values of the rescaled time $u = J^{1/2} t$. The initial state was a result of an evolution from $J = 0$ to $J = 0.01$ with $\tau_Q = 512$. Than it was allowed to thermalize. The expected long range order is not reached in finite time. However, the short range correlations do thermalize quickly, and so do the local observables, such as the energy. Thus, we can apply the equipartition principle. }
\label{therm3D}
\end{center}
\end{figure}

Since the parameter $J$ could be eliminated from Eq. (\ref{Josephson}) by introducing a rescaled time variable $u=J^{1/2}t$, the equations have a characteristic time-scale 
\begin{equation}
\tau ~\simeq~ J^{-1/2}~.  
\label{tau}
\end{equation}
This timescale is a relaxation time towards thermal equilibrium in the Josephson regime. In one dimension the system has finite correlation length and quickly reaches its equilibrium. The thermalization for 1D is depicted in Fig.~\ref{therm1D}.  
For two and three dimensions the situation is different. In 2D the correlations of the system display quasi-long-range order and in 3D they are of a long range order. Therefore, the two- or three dimensional system cannot equilibrate in finite time. However, short range correlation functions can reach equilibrium in finite time, which is shown in Fig~\ref{therm3D}. The correlations will be discussed in more detail in section~\ref{correl}. 

Since the kinetic and potential energies are quadratic in their degrees of freedom, Eq.~(\ref{energyInJosephson}), we can apply the equipartition principle and write
\begin{eqnarray}\nonumber
E_{\rm kin} = \frac12 T V ~,\\
E_{\rm pot} = \frac12 T V ~,
\label{equipart}
\end{eqnarray}
where we introduced temperature $T$ which is a function of $J$ in the adiabatic quench ($\tau_Q \gg 1$) and $V = L^D$ is the volume of the system, i.e. the number of sites. Eqs.~(\ref{equipart}) are valid for the Josephson regime, $J \ll 1$. The total energy is, obviously, $E = E_{\rm kin} + E_{\rm pot} = T V$

For adiabatic processes, where the system remains in equilibrium for every instant of time, both $T(t)$ and $J(t)$ are functions of time $t$. Hence we can write
\begin{equation}
\frac{d}{dt} E = \frac{dT}{dt}\, V ~,
\label{dedt.one}
\end{equation}
where we assume the energy to be a function of a time dependent temperature and 
\begin{equation}
\frac{d}{dt} E = \frac{dJ}{dt}\, E_{\rm kin}\, J^{-1} = \frac{dJ}{dt}\, \frac{TV}{2J} ~.
\label{dedt.two}
\end{equation}
Equating~(\ref{dedt.one}) with~(\ref{dedt.two}) and solving for $T(J)$ we get the equation characterizing the thermodynamic process induced by driving $J$:
\begin{equation}
T = A\, \sqrt{J} ~,
\label{adiabate}
\end{equation}
which we will call the adiabate equation. Again, it is valid for an adiabatic processes in the Josephson regime. 

\section{Excitation energy in the Josephson regime}

Having the adiabate equation~(\ref{adiabate}), we may now attempt to predict the scalings of energies~(\ref{energyInJosephson}) using the impulse-adiabatic approximation introduced in section~\ref{impAd}. The transition rate is
\begin{equation}
\frac{\dot{J}}{J} = \frac{1}{t}
\end{equation}
and the relaxation rate of the system
\begin{equation}
\tau^{-1} = \sqrt{J} = \left( \frac{t}{\tau_Q} \right)^{1/2} ~.
\end{equation}
The cross-over takes place at the instant of time $\hat{t}$ (corresponding to the parameter value $\hat{J}$) where these two rates become equal. Thus, 
\begin{equation}
\hat{t}^{-3/2} \simeq \tau_Q^{-1/2}
\end{equation}
or
\begin{equation}
\hat{J} \simeq \tau_Q^{-2/3} ~.
\label{jhat}
\end{equation}
The idea behind the impulse-adiabatic scenario is that the initial state remains frozen throughout the impulse stage of the evolution. Thus, the adiabatic process begins at $\hat{J}$ with random phases $\theta_{\bf i}$, 
\begin{equation}
E_{\rm kin} \big|_{\hat{J}} \approx \hat{J}\, \overline{\sum_{\bf i} \nabla \theta_{\bf i} \cdot \nabla \theta_{\bf i}} \big|_{\theta {\rm -random}} = \frac{2 \pi^2}{3} \, \hat{J} V D ~.
\end{equation}
The above equation gives the initial temperature for the adiabatic evolution
\begin{equation}
T \big|_{\hat{t}} \simeq \frac{4 \pi^2}{3} \hat{J} D ~.
\end{equation}
Comparing this formula with the adiabate equation allows us to determine the coefficient $A = \frac{4 \pi^2}{3} \sqrt{\hat{J}} D$. The adiabate equation becomes
\begin{equation}
T \simeq \frac{4 \pi^2}{3} D \sqrt{\hat{J}\, J} \simeq \sqrt{J} \tau_Q^{-1/3} ~,
\end{equation}
where in the last step we used the scaling for $\hat{J}$ (Eq.~(\ref{jhat})). As a consequence, we receive the algebraic scaling for kinetic and potential energy
\begin{equation}
E_{\rm kin} \simeq J^{1/2} \tau_Q^{-1/3} V ~, \quad
E_{\rm pot} \simeq J^{1/2} \tau_Q^{-1/3} V
\label{onethird}
\end{equation}
in the Josephson regime. 

\begin{figure}[!ht]
\begin{center}
\includegraphics[width=.9\textwidth,clip=true]{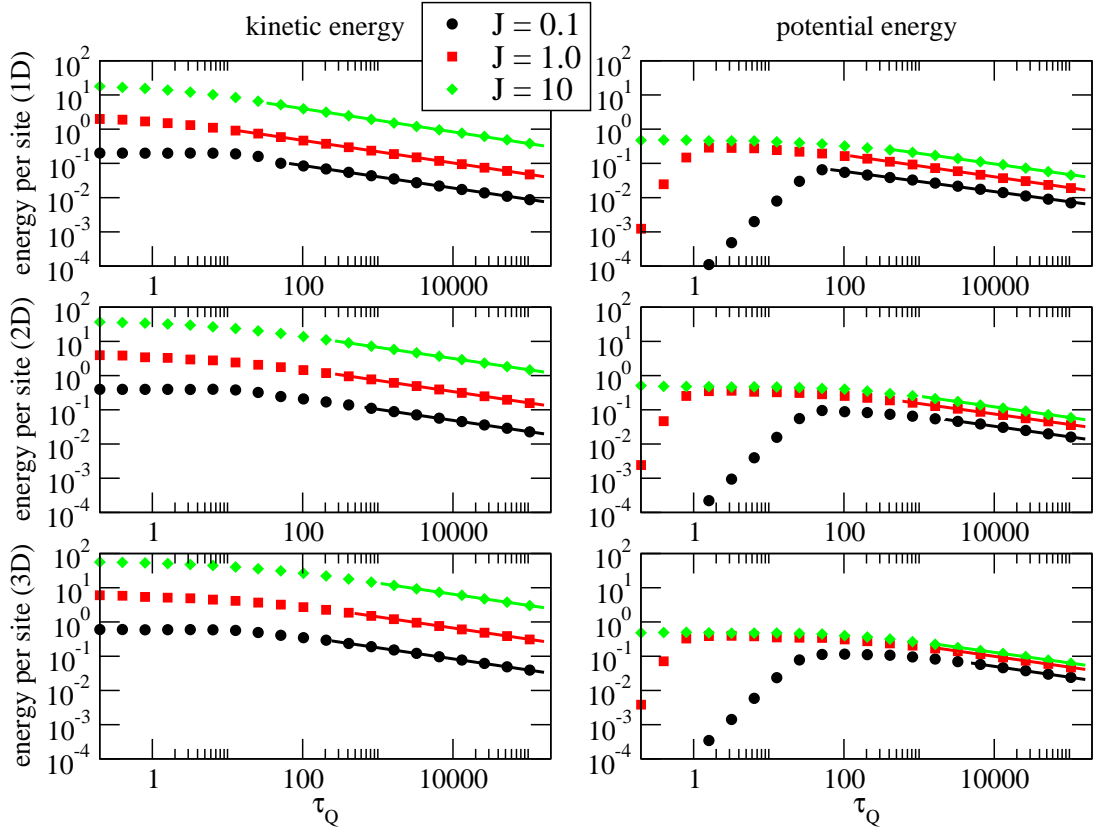}
\caption{The numerical evolution of the Eq.~(\ref{GPE}) leads to the kinetic and potential energies decay algebraically with $\tau_Q$. The exponent is numerically equal to $-\frac13$, see Tab.~\ref{tabener}, in agreement with the prediction~(\ref{onethird}). The lattice sizes were $L = 4096$, $256$, $128$ for 1, 2 and 3 dimensions respectively.}
\label{scalings}
\end{center}
\end{figure}

The predicted scaling agrees perfectly with the numerical simulation of the full Gross-Pitaevskii equation,~(\ref{GPE}), where we also get algebraic scalings with the exponent numerically close to $1/3$. The calculated $E_{\rm kin}$ and $E_{\rm pot}$ are presented in Fig.~\ref{scalings} and the fitted exponents are collected in Table~\ref{tabener}. 
\begin{table}[h!]
\begin{tabular}{ccccccc}
\hline \hline
$J$ & 
$\langle E_{\rm kin}^{\rm 1D}\rangle$ & 
$\langle E_{\rm kin}^{\rm 2D}\rangle$ & 
$\langle E_{\rm kin}^{\rm 3D}\rangle$ & 
$\langle E_{\rm pot}^{\rm 1D}\rangle$ & 
$\langle E_{\rm pot}^{\rm 2D}\rangle$ & 
$\langle E_{\rm pot}^{\rm 3D}\rangle$ 
\\ \hline
$0.1$  & $0.33$ & $0.33$ & $0.33$ & $0.33$ & $0.30$ & $0.30$\\
$1.0$  & $0.33$ & $0.33$ & $0.33$ & $0.33$ & $0.31$ & $0.31$\\
$10.0$ & $0.33$ & $0.33$ & $0.33$ & $0.32$ & $0.31$ & $0.31$\\
\hline \hline
\end{tabular}
\caption{ 
The best fits to $\alpha$ in $\langle E_{\rm kin/pot}\rangle\sim\tau_Q^{-\alpha}$ are consistent with $\alpha=1/3$. Lattice sizes as in Fig.~\ref{scalings}. 
}
\label{tabener}
\end{table}

\section{Excitation energy in the Rabi regime}

The data presented in Fig.~\ref{scalings} suggest, that the $\tau_Q^{-1/3}$ scaling extends also to the Rabi regime, $J \gg 1$. This can also be explained in terms of the impulse-adiabatic scenario. Since, for $J \gg 1$, the kinetic energy dominates, $\langle E \rangle \approx \langle E_{\rm kin} \rangle$, and~(\ref{dedt.two}) becomes
\begin{equation}
\frac{d}{dt} \langle E_{\rm kin} \rangle = \frac{dJ}{dt} \frac{\langle E_{\rm kin} \rangle}{J} ~.
\end{equation}
This has a simple solution $\langle E_{\rm kin} \rangle = B J$, where $B$ is a proportionality constant. The crossover from the Josephson to the Rabi regime is at $J \approx 1$ and thus we can determine the constant $B$:
\begin{equation}
\langle E_{\rm kin} \rangle \Big|_{J \approx 1} = J^{1/2} \tau_Q^{-1/3} V \Big|_{J \approx 1} = \tau_Q^{-1/3} V = B ~.
\end{equation}
Hence, in the Rabi regime, $\langle E \rangle \approx \langle E_{\rm kin} \rangle \approx J \tau_Q^{-1/3} V$. This is true, as long as the crossover takes place in the Josephson regime, which according to $\hat{J} \sim \tau_Q^{-2/3}$, happens for quenches slow enough. 

For fast quenches, the impulse-adiabatic crossover takes place in the Rabi regime. The thermalization time is set by the strength of the nonlinearity in~(\ref{GPE}) ($\tau \sim 1$) and, consequently, 
\begin{equation}
\frac1J \frac{dJ}{dt} \Bigg|_{\hat{J}} = \frac{1}{\hat{t}} \approx \tau^{-1} \sim 1 ~.
\end{equation}
Therefore, $\hat{t} \approx 1$. Since at $J$ phases remain random, $\langle E_{\rm kin} \rangle \Big|_{\hat{J}} \simeq \hat{J} V$. Finally, 
\begin{equation}
\langle E_{\rm kin} \rangle = \frac{J}{\hat{J}}\, \langle E_{\rm kin} \rangle \Big|_{\hat{J}} = \frac{2 \pi^2}{3} D J V ~.
\end{equation}
For fast quenches, energy does not depend on $\tau_Q$, consistently with the data from Fig.~\ref{scalings}. 

\section{Excitation energy for the system in a harmonic trap}

In order to make our analysis even more realistic and corresponding to the experimental conditions, we place our system in an external harmonic trap, i.e. quadratic potential. This amounts to setting the $V_{\bf i}$ in the Bose-Hubbard Hamiltonian~(\ref{BH}) to
\begin{equation}
V_{\bf i} = \frac12 \omega^2 {\bf i}^2 ~,
\end{equation}
where $\omega$ is the frequency of the trap, which sets its width, and we assume ${\bf i} = 0$ at the center of the lattice. Now, the Gross-Pitaevskii equation becomes
\begin{equation}
i\, \partial_t \phi_{\bf i} = 
-J \nabla^2\phi_{\bf i} + 
\left(|\phi_{\bf i}|^2-1\right)\phi_{\bf i} + \frac12 \omega^2 {\bf i}^2 \phi_{\bf i} \label{GPEinTrap}~,
\end{equation}
where the initial conditions are again random phases $\theta_{\bf i}$ and the density is distributed according to
\begin{equation}
|\phi_{\bf i}(0)|^2 = \frac{\omega^2}{2} (R^2_{\rm TF} - {\bf i}^2) ~,
\end{equation}
for ${\bf i}^2 < R_{\rm TF}^2$ and $|\phi_{\bf i}(0)|^2 = 0$ for ${\bf i}^2 \geq R_{\rm TF}^2$. This is an equilibrium solution to~(\ref{GPEinTrap}) for $J = 0$. In the above distribution $R_{\rm TF}$ is a radius (Thomas-Fermi radius) of a sphere inside which the distribution is non-zero. We choose $\omega^2 = \frac{2}{R^2_{\rm TF}}$, so that in the center of the trap initial field density is equal to $1$, $|\phi(0)|^2 = 1$. This makes comparison with the uniform case easier. 

In order to examine scaling of excitation energy in case of the presence of the harmonic trap, we have to subtract the ground state energy from the total energy calculated in the numerical simulation. The ground state energy can be computed as a solution to 
\begin{equation}
\frac{\delta E[\phi]}{\delta \phi_{\bf i}} = 0 ~,
\label{trapMinimalization}
\end{equation}
which minimizes the energy functional
\begin{equation}
E[\phi_{\bf i}] = \sum_{\bf i} \left\{ J \nabla \phi_{\bf i}^* \nabla \phi_{\bf i} + \frac12 \phi_{\bf i}^* \phi_{\bf i}^* \phi_{\bf i} \phi_{\bf i} + \frac{\omega^2}{2} {\bf i}^2 \phi_{\bf i}^* \phi_{\bf i} \right\} ~,
\label{energyFunctional}
\end{equation}
subject to the constraint
\begin{equation}
\sum_{\bf i} |\phi_{\bf i}|^2 = {\rm const.} ~,
\end{equation}
i.e. the norm should be kept constant throughout the evolution. The minimalization of~(\ref{energyFunctional}) can be done by evolving the following equation,
\begin{equation}
\frac{\partial \phi_{\bf i}}{\partial t} = - \frac{\delta E[\phi]}{\delta \phi_{\bf i}} ~,
\end{equation}
until $\phi_{\bf i}$ reaches the steady state, which corresponds to the solution of~(\ref{trapMinimalization}) and therefore is the ground state. 
\begin{figure}[!ht]
\begin{center}
\includegraphics[width=.9\textwidth,clip=true]{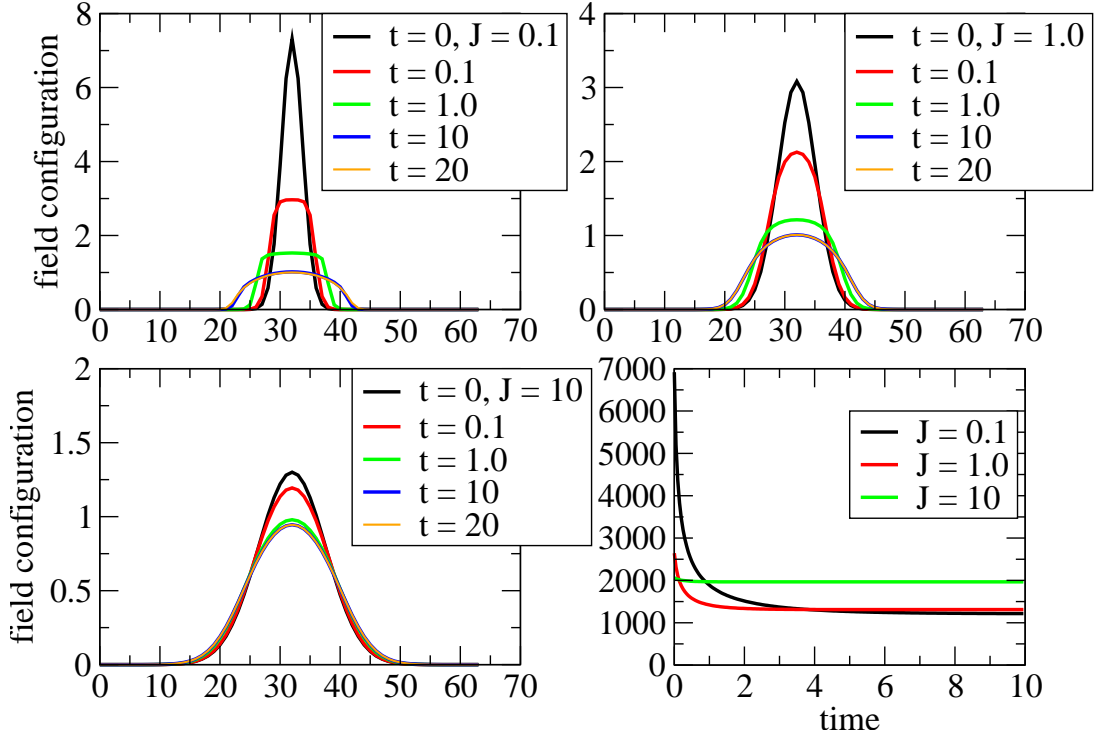}
\caption{Relaxation to a ground state of the energy functional $E[\phi_{\bf i}] = \sum_{\bf i} \left\{ J \nabla \phi_{\bf i}^* \nabla \phi_{\bf i} + \frac12 \phi_{\bf i}^* \phi_{\bf i}^* \phi_{\bf i} \phi_{\bf i} + \frac{\omega^2}{2} {\bf i}^2 \phi_{\bf i}^* \phi_{\bf i} \right\}$ according 
to the evolution of the equation $\frac{\partial \phi_{\bf i}}{\partial t} = - \frac{\delta E[\phi]}{\delta \phi_{\bf i}}$. Three panels represent subsequent field configuration
as it converges to the energy minimum and the bottom right panel shows the corresponding convergence of the ground state energy. The initial configuration was an educated guess: we assumed 
it is proportional to the ground state of the equation in the $J \gg 1$, ie. the harmonic oscillator equation. The larger $J$ the final ground state has more resemblance to the Gauss function 
as the harmonic oscillator approximation becomes more accurate. }
\end{center}
\end{figure}

\begin{figure}[!ht]
\begin{center}
\includegraphics[width=.6\textwidth,clip=true]{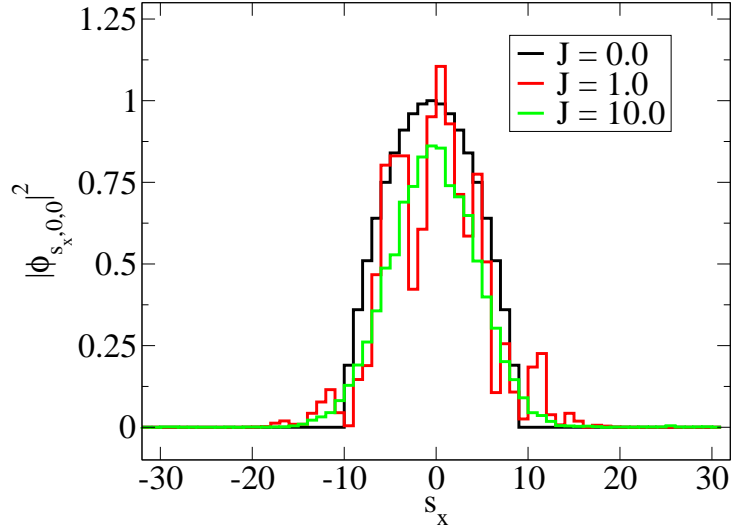}
\caption{The evolution of the cloud of atoms in a harmonic trap. The figure shows a cross-section through the center of a 3D lattice. The initial Thomas-Fermi profile at $J = 0$ expands and for large $J$ (here $J = 10$) tends to a Gaussian. The simulation for $\tau_Q = 102.4$. }
\label{cloud}
\end{center}
\end{figure}

The evolving cloud in the trap is initially centered in the form of a Thomas-Fermi distribution, but subsequently expands and asymptotically tends to the Gaussian distribution for large $J$. This is depicted in the Fig~\ref{cloud}. 

\begin{figure}[!ht]
\begin{center}
\includegraphics[width=.6\textwidth,clip=true]{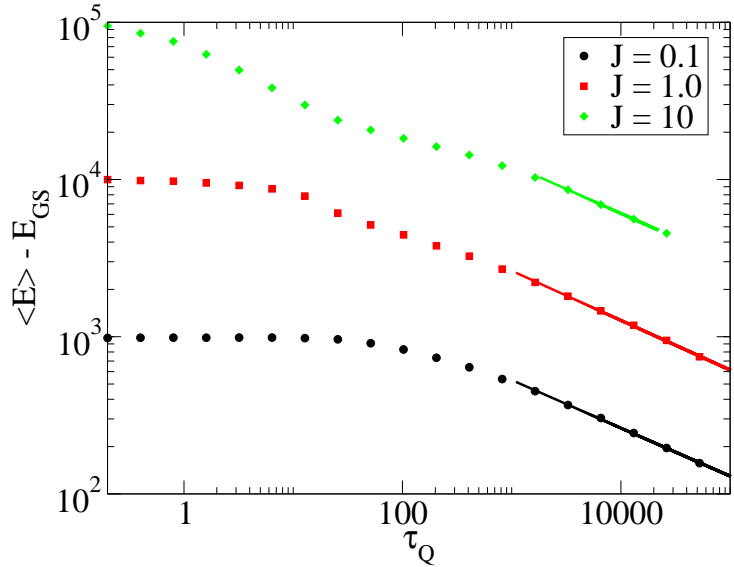}
\caption{The excitation energy $\langle E \rangle - E_{\rm GS}$ for the atoms in a trap. The energy scales algebraically with $\tau_Q$ and the powers are $0.32$ for $J = 1.0$ and $0.31$ for $J = 0.1$ and $J = 10.0$. The initial radius of the Thomas-Fermi profile was $R = 10$. }
\label{trapExcit}
\end{center}
\end{figure}

The numerical simulation of~(\ref{GPEinTrap}) indicates, that as in the uniform case, the excitation energy decays like $\tau_Q^{-1/3}$ with the quench rate. This data is shown in the Fig.~\ref{trapExcit}.  
The scaling obtained for a uniform system survives also in case of a trap, because the kinetic energy accumulated in random phase is much larger than the energy related to the localization of the cloud in the center of the trap. 

\section{Correlations and Vortices}
\label{correl}

\begin{figure}[!ht]
\begin{center}
\includegraphics[width=.6\textwidth,clip=true]{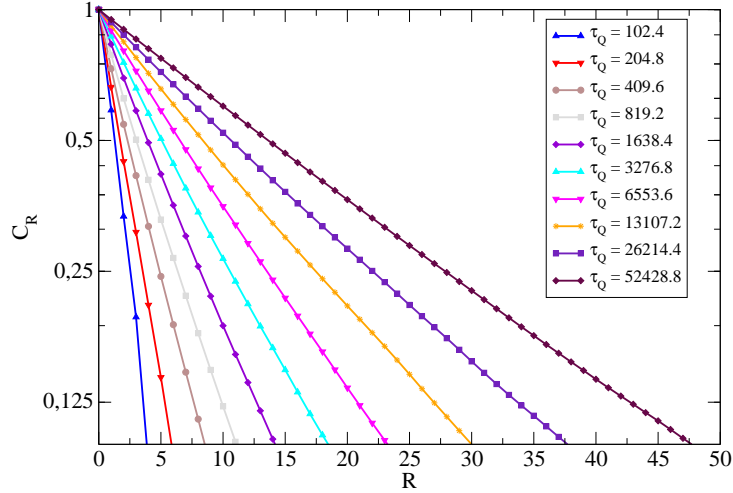}
\caption{The correlation functions in 1D at $J = 0.1$ are exponential, with the correlation length numerically scaling as $\xi \sim \tau_Q^{-1/3}$, consistently with~(\ref{length}). }
\label{correl1D}
\end{center}
\end{figure}

\begin{figure}[!ht]
\begin{center}
\includegraphics[width=.8\textwidth,clip=true]{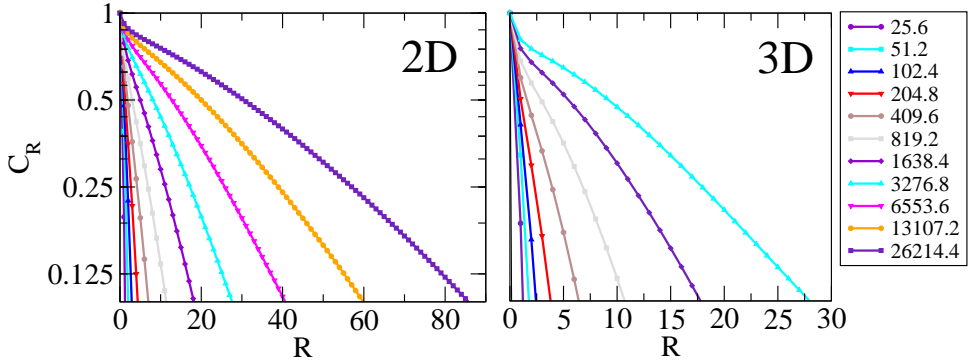}
\caption{Correlations at $J = 0.1$ in 2 and 3D respectively. The correlations have a finite range due to a limited pace of their growth. The local degrees of freedom have equilibrated, but at large scales the (quasi-)long-range order cannot be reached in finite time. }
\label{correl23D}
\end{center}
\end{figure}

In one dimension, the correlation length is finite, i.e. the correlation function is exponential, see derivation in~\cite{dtz}, 
\begin{equation}
C_R = \frac1n \langle a_{\bf i}^\dagger a_{\bf i} \rangle = e^{-\frac{R}{\xi}} ~,
\end{equation}
with $\xi$ depending on $J$ and $\tau_Q$ through
\begin{equation}
\xi = \frac{4J}{T} \simeq J^{1/2} \tau_Q^{1/3} ~.
\label{length}
\end{equation}
With growing $\tau_Q$ the correlation length decays like $\tau_Q^{-1/3}$, in agreement with the numerical data, presented in Fig.~\ref{correl1D}. 
This is not the case, however, in 2 and 3D, where there is a (quasi-)long-range order. Nevertheless, due to the finite rate, with which the system is driven, the correlations also spread with a finite rate and their range is limited, see Fig.~\ref{correl23D}. 

\begin{figure}[!ht]
\begin{center}
\includegraphics[width=.99\textwidth,clip=true]{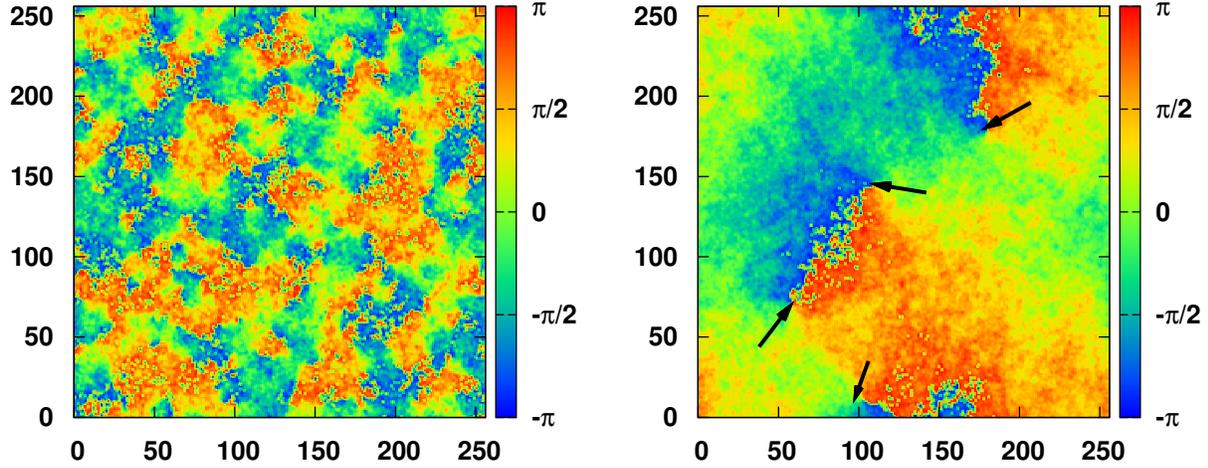}
\caption{The phase $\theta_{\bf s}$ for a 2D $256 \times 256$ system. The top panel shows the phase at $J = 0.1$ after a relatively fast quench with $\tau_Q = 1638.4$. In this case the domains that formed are small, as is the resulting correlation length $\xi \sim \tau_Q^{1/3}$. The bottom panel shows the situation after a quench with $\tau_Q = 52428.8$. The correlation length is much larger, the domains more prominent and the resulting vortices clearly visible (pointed with arrows). }
\label{vort}
\end{center}
\end{figure}

The finite rate of the correlation build-up results in the domain formation and appearance of topological vortices. Such vortices, in case of a 2D system, are presented in Fig.~\ref{vort}. For fast quenches (left panel), the there is little time for the correlations to spread, so the resulting pattern contains small domains and many vortices. When the system is allowed to cool down during a slow quench (right panel), the domains are larger and there are fewer, but more pronounced vortices. 



\section{Conclusion}

The physics beyond the linear quench from the Mott insulator phase to the superfluid turns out to be well approximated by the impulse-adiabatic scenario, where the phases remain frozen at random values, as for the initial state, but at some point the evolution becomes adiabatic and the system is allowed to thermalize. In 2 and 3 dimensions, however, the finite rate of the transition does prevent the (quasi-)long-range order to fully develop, and the resulting correlations have finite range. This is spectacularly manifested by the formation of vortices in phase. The simple picture of the impulse-adiabatic approximation allows for derivation of the algebraic decay of energies with the quench time $\tau_Q$, i.e. $E_{\rm kin/pot} \sim \tau_Q^{-1/3}$. This result is reproduced also when the system is subject to the external harmonic potential. This is in agreement with the experimental data obtained in~\cite{deMarco1}. 

\ack

This work was supported in part by the NCN grant 2011/01/B/ST3/00512 (JD and MT) and the PL-Grid Infrastructure (MT).

\end{document}